\documentstyle{article}

\author{Zhen Wang\\
Physics Department, LiaoNing Normal University, DaLian 116029, P. R. China}
\title{Measurement and Its Mathematical Scale
}
\input tcilatex

\QQQ{Description}{
}

\QQQ{Subject}{
02.10.By   03.65.Bz
}

\QQQ{Keywords}{
measurement, scale, Continuum Hypothesis
}

\QQQ{Language}{
American English
}

\begin{document}

\maketitle
\begin{abstract}
It is argued that every measurement is made in a certain scale. The scale in
which present measuments are made is called present scale which gives
present knowledge. Quantities at the limits to present measurement may be
observables in other scales. Cantor's series of infinites is used to
describe scales of measurement. Continuum Hypothesis and Schroedinger Cat
are discussed.
\end{abstract}

Measurement in quantum mechanics has been at the core of debate since
quantum mechanics was first born, because this problem is concerned with a
profound philosophical question: the relation between the outside world and
the observer. One of the many well-known paradoxes in quantum mechanics is
the Schroedinger Cat[1]. The question arises from this ideal experiment is :
does quantum mechanics describes true states in reality? In this paper we
are going to discuss the problem from the mathematical aspect of
measurement. We shall show that classification of states must corresponds to
measurement in a certain scale.

We know for any observable we have certain limit in its measurement. Ideal
measurement, which is infinitely accurate, is impossible in reality, just
like the case that a physical world could not exist without friction. Thus
mathematically there must be a smallest unit for any measurement. This
smallest unit represents the limit in the measurement, therefore it can not
be identified. Its nature is unknown to the observer. For these reasons, we
give it the name, uncertainty quantum, and denoted as $q$ . The interesting
point here is: although the uncertainty quantum composes a quantity, it can
not be figured out from the quantity itself. Otherwise it could not
represent the limit in the measurement. For examples, distance is composed
of space quantum, $q_l$ . This can be concluded from Zeno paradox[2]. Though
time is not an observable in physics, it is also in quantum. This can be
shown with a thought experiment[3].

Suppose we have an ideal, infinitely-high-rate model camera, with which we
can take infinite films in any short time interval. Of course we also have a
model projector which can show infinite number of films. Then let's aim at a
running dog. If time were not quantized, we would be able to take infinite
films of the running dog in any time interval. If then we show these
infinite number of films at the normal rate in cinema, the picture on the
screen would be motionless. Of course we can see a running dog if we show
the films at the same rate as we take them. But because of our persistence
of vision ({\em which may be a little bit different from person to person}),
we lose infinite information of the reality. We can not know what happens
within the interval of persistence of vision. That shows motion is the
direct result of our persistence of vision, which is the uncertainty quantum
of time for us human observer in this case. Thus time, and any quantity,
would be meaningless if it were not in quantum, the definite ignorance which
composes the observable.

So a quantity may be expressed as%
$$
F=\sum \,q_F\,\,\,\,\,\,\,\,\,\,\,\,\,\,\,\,\,\,\,\,\,\,\,\,\,\,\,(1) 
$$
where $q_F$ is the uncertainty quantum for $F$ . For any quantity to be
measurable, the uncertainty quantum must fulfill the infinitesimal
condition[4]:%
$$
F\pm q_F=F\;\;\;\;\;\;\;\;\;\;\;\;\;\;\;(2) 
$$
which may account for the genesis of continuity. That is to say, an
observable and its uncertainty quantum can not be identified at the same
time. Actually, the uncertainty quantum can not be identified {\em by
definition}, just like infinite. For the same reason, a certain infinite $%
\infty _F$ is defined. Thus we can express (1) as 
$$
F=q_F\cdot \infty _F\,\,\,\,\,\,\,\,\,\,\,\,\,\;\,\,\,\,\,\,\,\,(3) 
$$
In suitable unit, we can get the relation between uncertainty quantum and
the infinite:%
$$
q=\frac 1\infty \;\;\;\;\;\;\;\;\;\;\;\;\;\;\;\;\;\;\;(4) 
$$
It is evident that there is no absolute infinite and infinitesimal. All
infinite and infinitesimal are referred to some specific measurement. So the
uncertainty quantum and the infinite has defined a certain {\em scale}, in
which we do our measurement. If we adopt an opinion which is opposite to the
Anthropic Principle[5], i.e., we suppose no scale is special, then it is
quite straightforward to come to the point that all the infinities and
infinitesimals should be measurable in other scales, and symmetrically, all
observables can be infinitesimals or infinities in other scales.

Therefore it is necessary to make clear the scope of scale. Apparently this
means to study the structure beyond the infinite $\infty $ , or within the
uncertainty quantum $q$ . Thus our study is connected naturally with
Cantor's study of transcendental infinities. We know simple counting gives
only one infinite cardinal. That is, in the following series, there is only
one infinite cardinal $\aleph _0$:%
$$
1,2,\cdots \cdots ,\,\omega ,\,\omega +1\cdots \cdots ,\,2\omega ,\,\,\cdots
\cdots ,\,\omega ^2,\,\cdots \cdots ,\,\omega ^\omega ,\cdots \cdots
\;\;\;\;\;\;(5) 
$$
where $\omega $ is first infinite, the enumerable infinite. So just like
simple counting produces no new cardinals, new scales can not be defined by
extending present scale with any well-defined mathematical operation. We
simply take uncertainty quantum and infinite as the joining point of two
adjacent scales. This can be best expressed as:%
$$
\cdots \cdots ,\aleph _{-\alpha },\cdots \cdots ,\aleph _{-2}\,,\,\,\aleph
_{-1},\cdots \cdots 1,2,\,\cdots \cdots ,\aleph _0\,,\aleph _1,\cdots \cdots
,\,\aleph _\alpha ,\,\cdots \cdots \,\,\,\,\;\;\;\;\;(6) 
$$
in which $\aleph _{-1}$ is the uncertainty quantum for 1, 2, $\cdots \cdots $%
, $\aleph _{-2}$ is the uncertainty quantum for $\aleph _{-1}$, and so on.
It is obvious that observables in present scale 1, 2, $\cdots \cdots ,$ are
uncertainty quantum relative to quantities $\aleph _0$ in the next bigger
scale. This can be easily verified:%
$$
\aleph _0\pm n=\aleph _0\;,\;n\leq \aleph
_0\,\,\,\,\,\,\;\;\;\;\;\;\;\;\;\;\;\;\;\;\;\;\;\;(7) 
$$
Thus \{$\aleph _{-1},\aleph _0\}$ defines the {\em present scale} for
present measurements, in which $\aleph _{-1}$ is the uncertainty quantum, $%
\aleph _0$ is the infinite limit. Apparently the present scale represents
our knowledge, while other scales represent our ignorance. We show in other
work[3] that the uncertainty quantum contains all information in the present
scales. In fact all scales in the complement of the present scale are
connected, i.e., all our ignorance is connected.

Such classification of scales is necessary in dealing with problems
involving infinite quantities, because some infinite quantities may be in
different scales. If a problem is not in present scale, it has to be
transformed into the present scale, since we only have theories for problems
in present scale. The Continuum Hypothesis (CH) is an example[6]. In 1878
Cantor presumed that the weight of the real set is the second cardinal,
which was later expressed as 
$$
2^{\aleph _0}=\aleph _1\;\;\;\;\;\;\;\;\;\;\;\;\;\;\;\;\;(8)\;\; 
$$
Apparently the quantities in this problem are not in the present scale. To
solve the problem, we have either to establish theory dealing with problem
in other scale, or to transformed the problem into the present scale, i.e.
to change to the next bigger scale. The latter seems easier since the
arithmetics in this problem is very simple: it involves only multiplication.
From%
$$
\begin{array}{c}
n_1n_2\aleph _0=n_1+n_2+\aleph _0=\aleph _0,\;\;\;n_1,n_2\leq \aleph
_0\;\;\;\;\;\;\;\;\;\,\,\;\;(9) \\ 
\aleph _\alpha +\aleph _\beta =\aleph _\alpha \cdot \aleph _\beta =\max
(\aleph _\alpha \,,\aleph _\beta )\, 
\end{array}
$$
we can see that : {\em Multiplication would change to summation when it gets
into next bigger scale from present scale.} With this preposition we can
have new insight to CH problem.

If transformed to the next bigger scale, the CH would change to 
$$
2+2+\cdots \cdots =\aleph _0\,\;\;\;\;\;\;\;\;\;\;\;\;\;\;\;\;\;\;(10) 
$$
This is apparently true. One question may be raised: does the problem still
preserve its original significance after such transformation? We try to look
at the problem from a perspective outside mathematics. CH deals with an
unenumerable world and calls for new theory. Of course we live in an
enumerable world, in which we can count or enumerate (which is also a form
of measurement). Therefore the new theory required should bridge the
enumerable and the unenumerable world. In other words, CH is unprovable in
nature because it deals with something in the world of limitation to our
present knowledge. We can {\em NOT} prove it while preserving present
knowledge, since the world of limitation is exactly produced by present
knowledge. That is the significance of CH.

Thus our knowledge is expressed with well-defined states, which are
philosophically discontinuous so as to produce difference. But the
difference is composed of uncertainty quanta which can not be known in that
knowledge. That is to say, knowledge and ignorance are produced at the same
time. Actually, this is also true in the case of the Schroedinger Cat. In
defining the state ''dead'' and ''alive'', we produce the limitation about
the state ''dead'' and ''alive''. For an outside observer, ''unknown'' is
usual state in reality, just like the states ''dead'' and ''alive''. Here
the ''dead'' ,''alive'' and ''unknown'' are defined with the state of a
radio-active atom, rather than medical means. This is consistent with
quantum mechanics as well as daily experience. For an inside observer, it
seems that he has better and more natural means to define ''dead '' and
''alive'': his ability to think. But such an idea of ''measurement'' is
questionable : could one find that he could not think? In fact, one can not
define a ''dead'' state for oneself. For him the states ''dead'' and
''unknown'' get mixed, so that the definition of the state ''alive'' becomes
questionable. Thus the ''unknown'' state is different for the inside and
outside observers.

When ignorance is different, knowledge must be different. After all, quantum
mechanics tells only probability, though it is mathematically complete. The
uncertainty connects profoundly with our classification of states. In a
sense, we may even say our measurement decides the result of the measurement
to some extent. Not only in quantum mechanics, this may also be true in
classical realm[7]. From the above discussion on mathematical scale, we can
see that counting as a measurement inevitably introduces uncertainty into
mathematics. If mathematics is doomed to have uncertainty, probability may
be all we can ask from physics.

REFERENCES:

[1]Bernard d'Espagnat, {\it Conceptual Foundations of Quantum Mechanics}
(Benjamin 1971)

[2]Rozsa Peter, {\it Playing with Infinity} (Bell 1961)

[3]Zhen Wang, quant-ph/9806071

[4]Zhen Wang, quant-ph/9807035

[5]H. Pagels, {\it The Sciences}, Vol. 25, no.2, 1985

[6] P. J. Cohen, {\it Set Theory and Continuum Hypothesis} (W. A. Benjamin
Inc., New York 1966)

[7]Zhen Wang, quant-ph/9804070

\end{document}